\documentclass[pra, amsfonts, amssymb, amsmath,doi,reprint, showkeys, twoside,nobibnotes,floatfix, nofootinbib]{revtex4-1}

\usepackage[colorinlistoftodos, color=green!40, prependcaption]{todonotes}
\usepackage{hyperref}

\usepackage{amsthm}
\usepackage{mathtools}
\usepackage{physics}
\usepackage{xcolor}
\usepackage{graphicx}
\usepackage{adjustbox}
\usepackage{placeins}
\usepackage[T1]{fontenc}
\usepackage{lipsum}
\usepackage{csquotes}
\usepackage{textcomp}

\usepackage{url}
\usepackage{graphicx}
\usepackage{amsmath}
\usepackage{bm}
\usepackage{braket}
\usepackage[]{algorithm2e}

\usepackage{tabularx}

\begin{document}

\title{\emph{Ab initio} relativistic treatment of the intercombination $a^3\Pi - X^1\Sigma^+$ Cameron system of the CO molecule$^\dag$}
\footnotetext{\dag~Dedicated to the memory of our colleague and teacher Dmitry Alexandrovich Varshalovich}

\author{Nikolai S. Mosyagin$^{1}$}
\author{Alexander V. Oleynichenko$^{1,2}$}
\author{Andrei Zaitsevskii$^{1,2}$}
\author{Artur V. Kudrin$^{2}$}
\author{Elena A. Pazyuk$^{2}$}
\author{Andrey V. Stolyarov$^{2}$}
\email{avstol@phys.chem.msu.ru}

\affiliation{$^{1}$Petersburg Nuclear Physics Institute named by B. P. Konstantinov of National Research Center ``Kurchatov Institute'', 188300, Leningradskaya oblast, Gatchina, mkr. Orlova roscha, 1, Russia}
\affiliation{$^{2}$Department of Chemistry, Lomonosov Moscow State University, 119991, Moscow, Leninskie gory 1/3, Russia}

\date{\today}

\begin{abstract}
The intercombination $a^3\Pi - X^1\Sigma^+$ Cameron system of carbon monoxide has been computationally studied in the framework of multi-reference Fock space coupled cluster method with the use of generalized relativistic pseudopotential model for the effective introducing the relativity in all-electron correlation treatment. The extremely weak $a^3\Pi_{\Omega=0^+,1} - X^1\Sigma^+$ transition probabilities and radiative lifetimes of the metastable $a^3\Pi$ state were calculated and compared with their previous theoretical and experimental counterparts. The impact of a presumable variation of the fine structure constant $\alpha=e^2/\hbar c$ on transition strength of the Cameron system has been numerically evaluated as well.
\end{abstract}

\keywords{relativistic calculation; electronic transition probabilities; the Cameron system; carbon monoxide; fundamental constants variation.}

\maketitle              

\section{Introduction}
\begin{figure}[t!]
\includegraphics[scale=0.4]{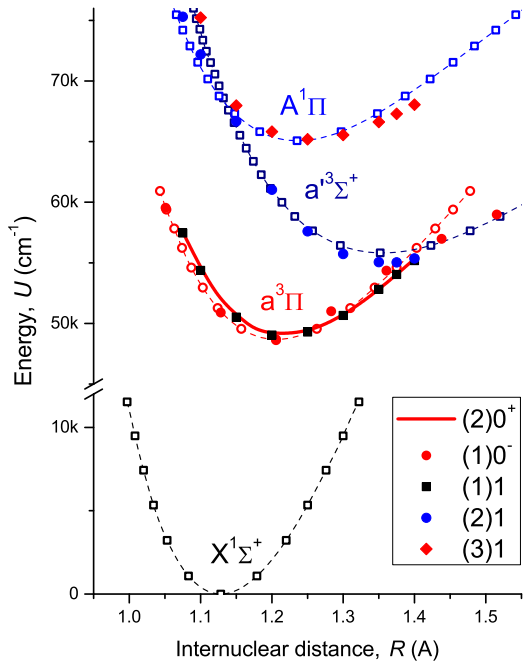}
\caption{Scheme of the low-lying electronic states of the CO molecule constructed from the empirical Rydberg-Klein-Rees potentials borrowed from Refs.~\cite{RKR_CO} (open symbols) and the present relativistic potential energy curves (solid symbols and line; $\Omega=0^+,1$ -- FS-RCC and $\Omega=0^-$ -- RCI) corresponding to the pure Hund's "\textbf{c}" coupling case.}\label{Fig_PEC}
\end{figure}

The carbon monoxide (CO) is the second most abundant diatomic molecule (after H$_2$) of the Universe, and, hence, the detailed data on its spectral properties are indispensably required to solve many fundamental and practical problems concerning the CO molecule. The $a^3\Pi - X^1\Sigma^+$ Cameron system of CO~\cite{Cameron1926}, lying in the ultraviolet region (170-270 nm), connects the ground singlet $X^1\Sigma^+$ state with the lowest excited $a^3\Pi$ state (see Fig.~\ref{Fig_PEC}). The upper triplet state is a metastable~\cite{Fournier, James, tauCO2007} since the spin-forbidden $a^3\Pi - X^1\Sigma^+$ electronic transition is extremely weak (its oscillator strength is only about 10$^{-7}$-10$^{-8}$~\cite{James, Minaev1995}). Nevertheless, the Cameron bands are well observed in both absorption and emission spectra due to the regular spin-orbit coupling of the $a^3\Pi$ state with the remote singlet states manifold~\cite{James_int}.

The rovibronic, fine and hyperfine structure as well as radiative, collisional, magnetic and electric properties of the ground and excited electronic states of the CO molecule have been comprehensively studied in a huge number of both experimental and theoretical works (see, for instance, the textbook~\cite{Field2004book} and references therein). However, it is not totally the case for the extremely weak $a^3\Pi - X^1\Sigma^+$ transition in spite of the a long time experimental efforts devoted to radiative lifetime determination of the $a^3\Pi_{\Omega^{\pm}=0,1,2}$ substates~\cite{Fournier, James, tauCO2007, tauCO1999, tauCO2000} as well as to intensity measurements for the band structure of the $a^3\Pi - X^1\Sigma^+$ transition~\cite{James_int}. The intercombination (triplet-singlet) Cameron system has never been studied under fully relativistic approximation combined with state-of-art electron correlation treatment. To the best of our knowledge, \emph{ab initio} studies of the $a^3\Pi - X^1\Sigma^+$ transition probabilities are limited so far by the multi-configurational self-consistent field (MCSCF) quadratic response approach~\cite{Minaev1995} or by using the spin-orbit coupling perturbation theory~\cite{Fournier, James, tauCO2007}.

Due to high sensitivity of the rovibronic energies of molecular lines to nuclear and electron masses, the high-redshift quasar absorption spectra of H$_2$ and CO molecules~\cite{Varshalovich, Ivanchik}  are basically used to probe a temporal variation of the proton-to-electron mass ratio ($m_p/m_e$)~\cite{Ubachs2016}. In particular, the Cameron bands of the CO molecule were identified in a quasar spectrum QSO 1556+3517 with the redshift $z=1.48$~\cite{Dubrovich1999}. The unprecedently accurate spectroscopic measurement on the $a^3\Pi - X^1\Sigma^+(0,0)$ band of various isotopomers of CO confirms the extreme sensitivity of electronic transitions involving nearly degenerate rovibronic levels for probing the variation of $m_p/m_e$ on the laboratory time scale~\cite{Ubachs2011}. Search for a presumable drift of the fine structure constant $\alpha=e^2/\hbar c$ is contrarily investigated via the atomic lines measurement~\cite{Murphy}. At the same time, the impact of the $\alpha$ change on the spin-orbit splitting and the intercombination $a^3\Pi - X^1\Sigma^+$ transition probabilities of CO should be apparently expected, thus, allowing to consider simultaneously both relativistic and mass effects. For the experimental verification of the possible drift of the fundamental constants the sensitivity coefficients for the wavelengths and intensities of the corresponding molecular transitions to variations of the parameters $m_p/m_e$ and $\alpha$ ($K^{\mu}$ and $K^{\alpha}$) are crucial~\cite{Meshkov2006}. The impact of scalar relativistic effects on the spectral characteristics of the isolated $X^1\Sigma^+$ state of the CO molecule has been recently studied~\cite{Konovalova2018}. The highly accurate $K^{\mu}$ and $K^{\alpha}$ estimates for most electronic transitions in CO is a non-trivial task which indispensably requires the quantum modeling beyond the conventional non-relativistic and adiabatic (Born-Oppenheimer) approximations~\cite{Bernath2005book}.

Therefore, the purposes of this work were to obtain the most reliable estimates for the very weak $a^3\Pi - X^1\Sigma^+$ transition probabilities and radiative lifetimes of the metastable $a^3\Pi$ state of the CO molecule as well as to investigate the sensitivity of transition strength of the CO Comeron system to a presumable drift of the fine structure constant, $\alpha$, on the cosmological time scale.

\section{Computational machinery}

The spin-forbidden $a^3\Pi_{\Omega=0^+,1} - X^1\Sigma^+$ transition dipole moments (TDMs) of the CO molecule, as functions of the internuclear distance, $R$, were calculated \emph{ab initio} by means of  three computational methods taking into account relativistic and electron-correlation effects in a different manner (see sections~\ref{GRPP}-\ref{MRCI} for details).

\emph{A priori} most accurate results were obtained in the framework of fully relativistic multi-reference relativistic Fock space coupled cluster calculations (FS-RCC) (section~\ref{FSRCC}) which utilized the advantages of the finite-field approach to transition property evaluation and generalized relativistic pseudo-potentials (GRPPs) to simulate relativistic structure of both C and O atoms. It should be noticed that the originally constructed GRPPs (section~\ref{GRPP}) allow one to treat the Coulomb interactions of \emph{all} electrons of the both atoms explicitly (so-called ``empty-core'' GRPPs). Under the conventional non-relativistic approximation (with the point nuclear model) the corresponding GRPP of the light atoms vanish.

The fully relativistic $a^3\Pi_{0^+,1} - X^1\Sigma^+$ TDM functions were also evaluated using the large scale multi-reference configuration interaction (MRCI) method employing the Dirac--Coulomb  electronic Hamiltonian with the incorporation of Gaunt interactions at the spinor generation stage and the so-called ``exact'' (X2C) transformation to the two-component picture~\cite{Ilias:07} (section ~\ref{RMRCI}). To monitor the MRCI energy and wavefunction convergence the various compositions of and active spaces were considered. The relativistic MRCI calculation (further referred as to RCI) performed with the DIRAC package~\cite{DIRAC:19} was found to be rather time consuming process. Moreover, its convergence strongly depends on the particular active space choice and is frequently affected by numerical instabilities.

Alternatively, the spin-orbit interaction between $a^3\Pi$ and the low-lying singlet states manifold of the CO molecule has been approximately accounted in a perturbation manner (section~\ref{MRCI}) using the \emph{scalar-state-interaction} (SSI) method, which is based on a diagonalization of the entire electronic Hamiltonian $\hat{H}_{sr}$+$\hat{H}_{so}$ built in a limited basis of the eigenfunctions of the scalar-relativistic Hamiltonian $\hat{H}_{sr}$. The required eigenvalues and eigenfunctions of the scalar-relativistic electronic Hamiltonian $\hat{H}_{sr}$ were obtained by means of the internally contracted MRCI method implemented in the MOLPRO software~\cite{MOLPRO2012}.

The resulting \emph{ab initio} $a^3\Pi_i - X^1\Sigma^+$ transition dipole moments, $d_{a^3\Pi_i - X}(R)$, and the corresponding difference potentials, $U_{a^3\Pi_i}(R) - U_X(R)$, were then applied for the radiative lifetime $\tau(\Omega^{\pm},v^{\prime},J^{\prime})$ estimates of the fine structure $\Omega=0,1,2$-components of the $a^3\Pi$ state as a function of vibrational $v^{\prime}$ and rotational $J^{\prime}$ quantum numbers. The required multi-component (non-adiabatic) vibrational wave functions of the $a^3\Pi_{\Omega^{\pm}}$ substates were obtained during rigorous close-coupled (CC) calculations (section~\ref{rtau}) which accounted explicitly for the spin-orbit splitting of the triplet state as well as the spin-rotational interaction between its $\Omega$-components~\cite{Klemperer}.

The dependence of the electronic excitation energies, $\Delta U_{\Omega-X}(R)=U_\Omega (R) - U_{X0^+}(R)$, and the corresponding electric dipole moments, $d_{\Omega-X}(R)$, for two lowest $\Delta \Omega =\pm 1$ relativistic transitions in the CO molecule, $(1,2)\Omega-X0^+$, on the variation of the fine structure parameter $\alpha$, is usually expressed in terms of the dimensionless sensitivity coefficients
\begin{equation}\label{Kalpha}
K^{\alpha}_{f}=\frac{\partial f}{\partial \alpha}\cdot\frac{\alpha}{f}
=\frac{\partial \ln (f)}{\partial \ln (\alpha)},
\end{equation}
where $f$ stands for the excitation energy ($\Delta U$) or absolute value of transition moment ($d$). These quantities were evaluated within the finite-difference approximation, changing the speed of light in the FS-RCC calculations from the standard value $c=137.03599911$~a.u. to $c_{-}=c/\sqrt{1.1}$ and $c_{+}=c/\sqrt{0.9}$. This was performed \emph{via} constructing two additional pseudopotentials for each atom with the modified speeds of light, $c_{-}$ and $c_{+}$.

\subsection{Empty-core GRPP for light elements}\label{GRPP}

The scheme of the generalized relativistic core effective potential generation developed for heavier elements in Refs.~\cite{Tupitsyn:95, Mosyagin:97, Petrov:04b, Mosyagin:06amin} was applied here with minor modifications. The theoretical background of the latter approach can be found in Ref.~\cite{Titov:99} and the latest versions are reviewed in Refs.~\cite{Mosyagin:16, Mosyagin:17, Mosyagin:20a}. In particular, the fully-relativistic Dirac-Fock-Breit (DFB) calculation was used to obtain the four-component spinors and their energies for the model state. Then, non-relativistic-type Hartree-Fock equations in the {\it jj}-coupling scheme were inverted to derive the GRPP components (potentials). Thus, these components effectively took into account for the relativistic effects.

The question immediately arises about the accuracy of the resulting model. The transition energies between low-lying states of the C atom and its cation calculated with the help of different methods are listed in Table~\ref{C_trans}. The Dirac-Fock-Breit results (with the Fermi nuclear charge distribution model) tabulated in the second column are used as the reference values. The errors of their reproducing in other calculations are listed in the following columns. The distinctions due to the point nuclear model and perturbative accounting for Breit interactions are negligible in comparison with the relativistic effects and are not presented in the Table (they are just zero at the level of the accuracy used in the Table). The errors of the Dirac-Fock (DF) calculations without Breit interactions and with the speed of light enlarged 1000 times, i.e. with the effect of relativity practically switched off (``HF'' in the Table~\ref{C_trans}) demonstrate the contributions of Breit and relativistic effects which are going to be simulated by the GRPP. One can see that the errors of the GRPP model (the 5th column) are more than one order of magnitude lower than the contributions of the relativistic effects (in the 4th column) and a few times lower than the contribution of Breit effects (the 3th column). The constructed numerical potentials were replaced by their Gaussian approximations without any detectable loss of accuracy.

Finally, the errors of semilocal Valence and Core versions derived from the above (Full) GRPP by neglecting the difference between the potentials for the $1s$ and $2s$ spinors are listed in the two last columns. One can see that the Valence GRPP version (with the potentials optimized for reproducing $2s$ and $2p$ spinors) is still acceptable for simulating the relativistic effects. Being compatible with most codes for relativistic electronic structure modelling, this version provides a useful (and, at the time being, unique) tool offering the possibility of describing Breit interactions in molecular all-electron calculations. This seems of particular importance for light-element compounds where the Breit interaction frequently have non-negligible contribution in contrast to rather weak spin-dependent relativistic effects. It should be noted that the potential acting on the $s$-electrons in the Valence version was constructed for the $2s$ spinor with large component having a radial node (unlike $1s$ for the Core version).

\begin{table*}
\caption{Transition energies ($\Delta E$) between some relativistic terms and states averaged over the non-relativistic configurations of the C atom and its cation from numerical DFB calculations and the corresponding absolute errors of their reproducing in the different versions of DF and GRPP calculations. All the values are in cm$^{-1}$.}\label{C_trans}
\begin{tabular}{lrrrrrr}
\hline\hline
Configuration                          & DFB        & DF    & HF          &\multicolumn{3}{c}{GRPP}\\
\cline{5-7}
(Term)                                 &            &       &             & Full  & Val.\   & Core  \\
\hline
                                       & $\Delta E$ & Error & Error       & Error & Error   & Error \\
\hline
\multicolumn{7}{l}{Nonrel.aver.\ $1s^2 2s^2 2p^2 \rightarrow$}                                      \\
 $1s^2 2s^2 2p^1 3s^1                $ &      52415 &     4 &          47 &     1 &      -1 &    24 \\
 $1s^2 2s^2 2p^1                     $ &      80611 &     5 &          40 &     0 &      -1 &    20 \\
 $1s^2 2s^2      3s^1                $ &     195398 &    13 &         105 &     0 &      -5 &    55 \\
 $1s^2 2s^1 2p^3                     $ &      70858 &     7 &        -114 &     1 &       4 &   -57 \\
 $1s^2 2s^1 2p^2 3s^1                $ &     117676 &    10 &         -85 &     3 &       4 &   -43 \\
 $1s^2 2s^1 2p^2                     $ &     145447 &    11 &         -93 &     2 &       4 &   -47 \\
 $1s^2 2s^1 2p^1 3s^1                $ &     250149 &    19 &         -49 &     3 &       1 &   -24 \\
 $1s^2      2p^3                     $ &     233835 &    17 &        -234 &     1 &       8 &  -119 \\
\hline
\multicolumn{7}{l}{Rel.term $\ldots 2p_{1/2}^1 2p_{3/2}^1 (J=1) \rightarrow$}                       \\
 $\ldots            2p_{3/2}^2 (J=2) $ &       4288 &    14 &         -43 &     4 &       4 &     5 \\
 $\ldots 2p_{1/2}^1 2p_{3/2}^1 (J=2) $ &       8429 &     6 &           7 &     6 &       6 &     7 \\
 $\ldots 2p_{1/2}^2            (J=0) $ &      10463 &   -10 &          49 &     1 &       1 &     2 \\
 $\ldots            2p_{3/2}^2 (J=0) $ &      20726 &     9 &         -43 &    -2 &      -2 &     1 \\
\hline
\end{tabular}
\end{table*}

\subsection{Relativistic coupled cluster calculations}\label{FSRCC}

The Fock space relativistic coupled cluster \cite{Visscher:01} calculations for the CO molecule with the series of Valence empty-core GRPPs described above employed the standard \emph{aug-cc-pVQZ} basis sets~\cite{Dunning:89, Kendall:92} on both centers. The set of one-electron spinors and the Fermi vacuum state were obtained by solving the spin--orbit-coupled Kramers-restricted SCF equations for the ground-state configuration of the neutral CO molecule. Excited states were described within the one hole -- one particle ($1h1p$) Fock space sector; the ($1h1p$) model space was spanned by all singly excited configurations with a hole on one of 8 highest-energy occupied spinors and a particle on one of 24 lowest-energy virtual spinors. To prevent numerical instabilities in solving the FS-RCC amplitude equations, we used the adjustable denominator shift technique~\cite{Zaitsevskii:18a} in the form described in Ref.~\cite{Oleynichenko:20cpl} (``complex shift simulation''). In order to maintain core separability of results, no shifting has been applied in the vacuum ($0h0p$) sector. We also kept unshifted the denominators in the equations for the single de-excitation amplitudes in the ($1h1p$) sector. In the other sectors the shift amplitudes -0.15~\emph{a.u.} for single excitations and -0.30~\emph{a.u}.for double excitations with the attenuation parameter $m=3$ (see Eqs.(7--8) of Ref.~\cite{Oleynichenko:20cpl}) were assumed.

Transition dipole moments $D_{ij}$ between the electronic states $\psi_i,\;\psi_j$ of CO were calculated with the help of the finite-field technique~\cite{Zaitsevskii:18, Zaitsevskii:20t}, using the approximate relation
\begin{equation}\label{basicff}
|D_{ij}|\approx\Delta E_{ij}
\left|\langle \tilde{\psi}_i^{\perp\!\!\perp} \vert \frac{\partial}{\partial{}F}  \tilde{\psi}_j \rangle \right|^{1/2}
\left|\langle \tilde{\psi}_j^{\perp\!\!\perp} \vert \frac{\partial}{\partial{}F}  \tilde{\psi}_i \rangle \right|^{1/2}
\end{equation}
where $F$ is the strength of the applied uniform electric field, $\Delta E_{ij}$ stands for the absolute value of the $i\to j$ transition energy and $\{\tilde{\psi}_j\}$ are the (normalized) projections of the field-dependent electronic state wavefunctions $\psi_j$ onto the direct sum of the vacuum state and the ($1h1p$) model space, $\mathcal{L}^{(0h0p+1h1p)}$, which is constructed for $F=0$ and does not depend on $F$. The set of functions $\{\tilde{\psi}_i^{\perp\!\!\perp}\}$ is biorthogonal to the basis in $\mathcal{L}^{(0h0p+1h1p)}$ composed of $\{ \tilde{\psi}_j\}$:
\begin{equation}\label{biorth}
\tilde{\psi}_i^{\perp\!\!\perp}=S^{-1}{\psi}_i
\end{equation}
where $S$ is the overlap matrix: $S_{ij}= \langle \tilde{\psi}_i \vert  \tilde{\psi}_j \rangle$. The derivatives in Eq.(\ref{basicff}) are to be evaluated at $F=0$ and are estimated within the central finite-difference approximation, using the results of calculations with two different field strength values. In the present calculations the step of numerical differentiation was assumed to be equal to 0.00005 or 0.0001 \emph{a.u.}; the results obtained with these two step sizes were practically identical. Although this version of the finite-field technique is based exclusively on the analysis of model-space entities, the contributions to $D_{ij}$ from the parts of wavefunctions outside $\mathcal{L}^{(0h0p+1h1p)}$ are incorporated implicitly~\cite{Zaitsevskii:18, Zaitsevskii:98}.
It is worth noting that, in contrast with the complete-model-space case~\cite{Zaitsevskii:18}, the projections $\{\tilde{\psi}_{i}\}$ are not directly obtained as eigenfunctions of the FS-RCC effective Hamiltonian. However, these eigenfunction can be readily transformed into $\{\tilde{\psi}_{i}\}$, using the ``closed'' (i.\ e. acting within $\mathcal{L}^{(0h0p+1h1p)}$) part of the cluster operator~\cite{Zaitsevskii:20t}.

The required one-electron spinors and molecular integrals were evaluated using the DIRAC19 program package~\cite{DIRAC:19, DIRAC:20}; the FS-RCC calculations were carried out with the help of the EXP-T code~\cite{Oleynichenko:20a, Oleynichenko:20, website:expt}.

\subsection{Relativistic configuration interaction calculations}\label{RMRCI}

The relativistic MRCI calculations employed the uncontracted version of the all-electron correlation consistent \emph{aug-cc-pCVQZ} Dunning basis set~\cite{Woon:95}. At the first step of calculations, four-component calculations for the CO$^+$ cation were performed with the Dirac--Coulomb--Gaunt Hamiltonian to generate the set of one-electron spinors suitable for the description of excited electronic states of the neutral molecule. Then the four-component Dirac-Coulomb Hamiltonian was transformed into a two-component one, accurately reproducing the positive-energy spectrum of the original four-component Hamiltonian (X2C Hamiltonian~\cite{Ilias:07}) which was used to perform correlation calculations in the frame of the MRCI method~\cite{Fleig:03}. The correlations of 8 or 10 electrons of the CO molecule were taken into account, disregarding or accounting for the excitation of 2$s$ electrons of the O atom, whereas the excitations of 1$s$ electrons of C and O were always omitted. The configuration space included the excitations of up to two electrons from the complete-active-space spinors corresponding to the $4\sigma-6\sigma$, $1\pi,\;2\pi$ or $3\sigma-6\sigma$,  $1\pi,\;2\pi$ scalar MOs.

\subsection{Scalar-state-interaction calculations}\label{MRCI}

The calculation of individual electronic matrix elements of the spin-orbit operator $\hat{H}_{so}$ and transition dipole moment $\hat{d}$ between the scalar states, construction and diagonalization of the entire $\hat{H}_{sr}$ + $\hat{H}_{so}$ matrix were accomplished within the HLSMAT procedure~\cite{SO} incorporated into the MOLPRO package. Similar to the case of FS-RCC calculations described above, empty-core GRPPs and the all-electron~\emph{aug-cc-pV5Z} basis sets~\cite{Kendall:92} were used for both atoms. The initial Hartree-Fock molecular orbitals (MOs) were optimized within the state-averaged complete active space self-consistent field (SA-CASSCF) method~\cite{Werner85}, taking the (1-4)$^{1,3}\Sigma^+$ and (1-4)$^{1,3}\Pi$ electronic states with equal weights. The dynamic correlation was accounted for all 14 electrons of the molecule within the internally contracted MRCISD method~\cite{iMRCI}. The active space used in the SA-CASSCF calculation consisted of six $\sigma$ and four $\pi$ MOs while in the subsequent MRCISD calculation of (1-2)$^3\Pi$, (1-3)$^1\Sigma^+$ and (1-3)$^1\Pi$ states two inner core $\sigma$ orbitals were kept doubly occupied.

\subsection{Radiative lifetime calculations}\label{rtau}

The rovibronic eigenvalues and eigenfunctions for lowest rovibrational levels of the fine-structure components ($\Omega^{\pm}$, $v^{\prime}$, $J^{\prime}$) of the $a^3\Pi_{\Omega^{\pm}}$ state were obtained by solving the three close-coupled (CC) equations~\cite{DUO}:
\begin{eqnarray}\label{CC}
\left(- {\bf I}\frac{\hbar^2 d^2}{2\mu dR^2} + {\bf V}(R;\mu,J^{\prime}) - {\bf I}E^{CC}\right)\mathbf{\Phi}(R) = 0
\end{eqnarray}
with the conventional boundary $\phi_i(0)=\phi_i(\infty)=0$ and normalization $\sum_{i=1}P_i=1$ conditions, where $i\in[a^3\Pi_0;a^3\Pi_1;a^3\Pi_2]$ and $P_i=\langle\phi_i|\phi_i\rangle$ is the fractional $\Omega$-partition of the triplet state. The corresponding potential energy matrix ${\bf V}(R;\mu,J^{\prime})$ was taken in the form:
\begin{eqnarray}\label{Ham}
\langle ^{3}\Pi_{0}|H|^{3}\Pi_{0} \rangle & = & U_{a} - A^{so} + B(X+1),\nonumber \\
\langle ^{3}\Pi_{1}|H|^{3}\Pi_{1} \rangle & = & U_{a} + B(X+1),\nonumber \\
\langle ^{3}\Pi_{2}|H|^{3}\Pi_{2} \rangle & = & U_{a} + A^{so} + B(X-3),\nonumber \\
\langle ^{3}\Pi_{0}|H|^{3}\Pi_{1} \rangle & = & - B\sqrt{2X},\nonumber \\
\langle ^{3}\Pi_{1}|H|^{3}\Pi_{2} \rangle & = & - B\sqrt{2(X-2)},\nonumber
\end{eqnarray}
where $X=J(J+1)$, $B(R) = \hbar^{2}/2 \mu R^{2}$ while $U_a(R)$ is the empirical Rydberg-Klein-Rees (RKR) potential of the $a^{3}\Pi_1$ component~\cite{RKR_CO} and $A^{so}(R)$ is the spin-orbit splitting available in Ref.~\cite{James} as a polynomial function of $R$.

Since the $0^- - 0^+$ and $2 - 0^+$ transitions are strictly forbidden under relativistic approximation only the $a^3\Pi_{0^+,1} - X^1\Sigma^+_{0^+}$ transitions contribute to the radiative lifetime $\tau (\Omega^{\pm}, v^{\prime}, J^{\prime})$ of the $a^3\Pi$ state:

\begin{equation}\label{tau}
\frac{1}{\tau} = \frac{k}{2J^{\prime}+1}\sum_{v_X,J_X}\hspace{-0.5em}\nu^3_{a-X}
\left|\sum_{\Omega^{\prime}\in 0^+,1}\hspace{-0.7em}\langle \phi_{^3\Pi_{\Omega^{\prime}}}|d_{^3\Pi_{\Omega^{\prime}}-X}|v_X\rangle S^{\Omega^{\prime} \Omega_X}_{J^{\prime} J_X}\right|^2,
\end{equation}

where $k=2.02586232\times 10^{-6}$, $\nu_{a-X}=E^{CC}(\Omega^{\pm},J^{\prime})-E_X(v_X,J_X)$ is the transition wavenumber (in cm$^{-1}$), $d_{^3\Pi_{\Omega^{\prime}}-X}(R)$ are the transition dipole moments (in $a.u.$) and $S^{\Omega^{\prime} \Omega_X}_{J^{\prime} J_X}$ are the properly normalized dimensionless direction cosine matrix elements well-known in the analytical form ~\cite{Bernath2005book}. It is clear that the $a^3\Pi_{0^+} - X^1\Sigma^+_{0^+}$ transition does not make matter for the $\Omega^{-}$-components of the $a$-state.

To avoid a summation over rotational and vibrational levels of the ground state in Eq. (\ref{tau}), the approximate sum rule~\cite{Tellinghuisen:84,PupyshevCPL94} generalized for the CC case~\cite{Kiyoshima2003} can be used as
\begin{eqnarray}\label{tausum}
\frac{1}{\tau_a} \approx k\hspace{-0.5em} \sum_{\Omega^{\prime}\in 0^+,1}\hspace{-0.5em}\langle \phi_{^3\Pi_{\Omega^{\prime}}}|[\Delta U_{^3\Pi_{\Omega^{\prime}}-X}]^3\vert d_{^3\Pi_{\Omega^{\prime}}-X}\vert^2|\phi_{^3\Pi_{\Omega^{\prime}}}\rangle,
\end{eqnarray}
where $\Delta U_{^3\Pi_{\Omega^{\prime}}-X} = U_{a^3\Pi_{\Omega^{\prime}}} - U_X$. The advantage of the Eq.(\ref{tausum}) is that the phases of TDM functions and non-adiabatic wavefunctions can be undefined.

\section{Results and discussion}

\begin{figure}[t!]
\includegraphics[scale=0.4]{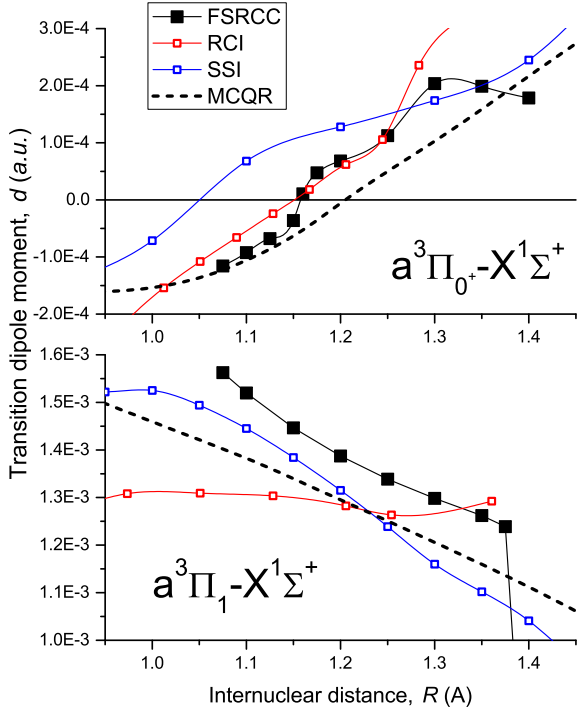}
\caption{The \emph{ab initio} $a^3\Pi_{0^+,1} - X^1\Sigma^+$ electronic transition dipole moments evaluated using the different computational schemes: FS-RCC - fully relativistic Fock space coupled cluster method combined with the finite-field (FF) approach and generalized relativistic effective core potentials (GRPPs), RCI - large scale fully relativistic multi-reference configuration interaction method, SSI - scalar-state-interacting calculation based on the GRPPs exploiting, MCQR - multi-configuration quadratic response approach as implemented in Ref.\cite{Minaev1995}.}\label{Fig_TDM}
\end{figure}

\begin{figure}[t!]
\includegraphics[scale=0.4]{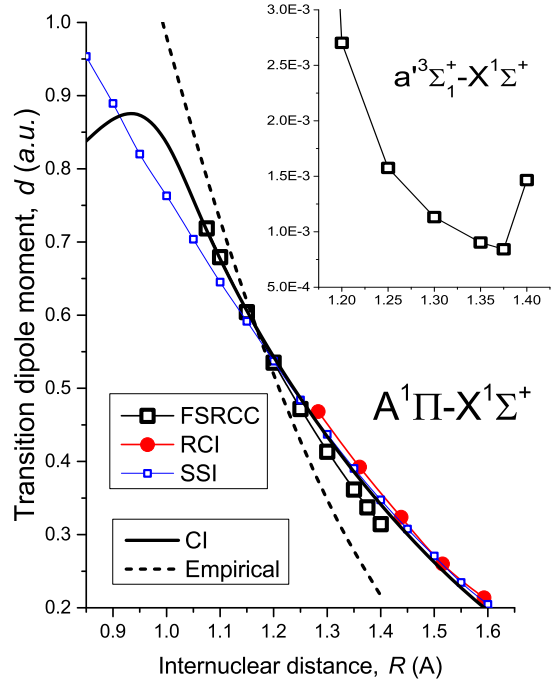}
\caption{The relativistic electronic transition moments obtained for the spin-allowed $A^1\Pi - X^1\Sigma^+$ transition in the framework of the above computational schemes: FS-RCC, RCI and SSI.
``CI'' stands for the non-relativistic MRCI calculations performed in Ref.~\cite{Kirby1989}; the empirical $A-X$ TDM function is borrowed from Ref.~\cite{Leon1988}. The inset presents the fragment of the spin-forbidden $a'^3\Sigma_1^+ - X^1\Sigma^+$ TDM function extracted from the $(2)1 - X0^+$ FS-RCC result for the $R\in [1.2, 1.4]$~(\AA) interval.}\label{Fig_AXTDM}
\end{figure}

The excitation energies, $\Delta U^{ab}_{\Omega-X}(R)$, and transition dipole moments, $d^{ab}_{\Omega-X}(R)$, obtained for the $(1,2)1-X0^+$ transitions by using the fully relativistic FS-RCC method are given in Table~\ref{TabTDM}, together with the relevant sensitivity coefficients, $K^{\alpha}_{f}$, evaluated according to Eq.(\ref{Kalpha}). The relativistic $(2)0^+-(1)0^+$ and $(1,2)1-(1)0^+$ TDM functions obtained in the framework of the computational schemes discussed above (FS-RCC, RCI and SSI) are depicted on Fig.~\ref{Fig_TDM} and Fig.~\ref{Fig_AXTDM}, where they are compared to each other and with previous theoretical~\cite{Kirby1989} and experimental~\cite{Leon1988} counterparts. Fragments of the resulting relativistic potential energy curves (PECs) $U_{\Omega}(R)$ are shown on Fig.~\ref{Fig_PEC} along with the empirical RKR potentials~\cite{RKR_CO} corresponding to the Hund's ``\textbf{a}'' coupling case~\cite{Field2004book}. The radiative lifetimes evaluated by Eq.(\ref{tausum}) for the particular rovibrational levels of the metastable $a^3\Pi_{0^{\pm},1,2}$ substates are compared with their experimental counterparts~\cite{Fournier, tauCO2007, tauCO1999, tauCO2000} in Table~\ref{Tabtau}. The complete tables of the \emph{ab initio} FS-RCC, RCI and SSI transition dipole moments and resulting radiative lifetimes are tabulated in machine-readable format in the Supplemented Material (SM).

It should be noted that the theoretical PECs were obtained for the excited $\Omega=0^{\pm},1$ states by adding the \emph{ab initio} calculated vertical excitation energies to the highly accurate empirical ground state potential $U^{emp}_X$ from Ref.~\cite{Meshkov2018}: $U_{\Omega}(R) = \Delta U^{ab}_{\Omega-X}(R) + U^{emp}_{X^1\Sigma^+}(R)$ (cf.~\cite{Zaitsevskii2005}). As expected the resulting relativistic PECs demonstrate the avoided crossing in the vicinity of the points $R_{A-a'}\approx 1.15$~(\AA) and $R_{a-a'}\approx 1.4$~(\AA) which correspond to the crossing of the singlet-triplet $A^1\Pi-a'^3\Sigma^+$ and triplet-triplet $a^3\Pi-a'^3\Sigma^+$ states, respectively. This effect is most clearly observed as abrupt changes of the relativistic $(1,2)1-X0^+$ TDM functions near the real crossing points of the relevant multiplet states (see Fig.~\ref{Fig_TDM} and Fig.~\ref{Fig_AXTDM}).

Overall good agreement is obviously observed among the present \emph{ab initio} TDM functions evaluated in the framework of the alternative computational schemes: FS-RCC, RCI and SSI. The results of the finite-field transition moment calculations within the FS-RCC method should be apparently considered as the most reliable ones since the radiative lifetimes based of the FS-RCC functions are found to be remarkably close to their most accurate experimental counterparts~\cite{tauCO2007} measured for (1$^+$,0,1) and (2$^+$,0,2) rovibrational levels of the $a^3\Pi$ state (see Table~\ref{Tabtau}). Furthermore, the FS-RCC $a^3\Pi_{\Omega=0^+,1} - X^1\Sigma^+$ TDM functions reproduce very well the distinguishable $\tau$-values measured for the different $\Omega^{\pm}$-components of the (0$^\pm$,3,2) level~\cite{tauCO2000}. At the same time, the present $\tau$ estimates for the (0$^+$,0,$J'$) level are significantly higher than the experimental 90 ms obtained for CO molecules trapped in solid neon matrices~\cite{Fournier}. This difference can be understood by bearing in mind the exponential decrease of the radiative lifetimes as $J'$-value increase, namely: $\tau$= 690, 250, 115 and 65 ms for $J'$= 0, 1, 2 and 3 levels, respectively.

The fully relativistic FS-RCC and RCI calculations generally confirm the validity of the previous MCQR TDM functions obtained for the $a^3\Pi_{\Omega=0^+,1} - X^1\Sigma^+$ transitions due to implementation of the multi-configuration quadratic response approach~\cite{Minaev1995} (see Fig.~\ref{Fig_TDM}). The relativistic electronic transition moments obtained for the spin-allowed $A^1\Pi - X^1\Sigma^+$ transition in the framework of the above FS-RCC, RCI and SSI computational schemes also agree very well with results of the conventional non-relativistic calculations~\cite{Kirby1989} and empirical data~\cite{Leon1988} (see Fig.~\ref{Fig_AXTDM}). It is interesting that the part of the relativistic FS-RCC $d_{(2)1 - (1)0^+}(R)$ function can be approximately assigned to the spin-forbidden $a'^3\Sigma_1^+ - X^1\Sigma^+$ transition.

The Table~\ref{TabTDM} clearly demonstrates that the sensitivity coefficients of TDM functions, $K^{\alpha}_d$, corresponding to spin-forbidden transitions are very close to 2 indeed, as it should be expected from the trivial argument of perturbation theory: $d\varpropto \hat{H}_{so}\varpropto \alpha^{2}$. At the same time, the $K^{\alpha}_d$-values for the spin-allowed transitions are 1000 times smaller. The sensitivity coefficients of the excitation energy, $K^{\alpha}_{\Delta U}$, corresponding to the Cameron system, is about $2\times 10^{-3}$ while for the higher $(2)1 - X0^+$ transition the relevant $K^{\alpha}_d$-values are about $2.5\div 3.5\times 10^{-4}$. All sensitivity coefficients are very smooth functions of internuclear distance except narrow regions in the vicinity of the avoided crossing points.

\begin{table*}
\caption{The excitation energies, $\Delta U_{\Omega-X}$  (in cm$^{-1}$), transition dipole moments, $d_{\Omega-X}(R)$ (in $a.u.$), and dimensionless sensitivity coefficients, $K^{\alpha}_d$ and $K^{\alpha}_{\Delta U}$, obtained for the relativistic $(1,2)1-X0^+$ transitions of the CO molecule as functions of internuclear distance $R$ (in~\AA) calculated at the FS-RCC level.}\label{TabTDM}
\begin{tabular}{ccccccccc}
\hline\hline
      & \multicolumn{4}{c}{$(1)1-X0^+$ transition} & \multicolumn{4}{c}{$(2)1-X0^+$ transition}\\
 $R$  & $\Delta U$ & $K^{\alpha}_{\Delta U}\!\times\! 10^{-3}$ & $d\!\times\! 10^{-3}$ & $K^{\alpha}_d$
      & $\Delta U$ & $K^{\alpha}_{\Delta U}\!\times\! 10^{-4}$ & $d\!\times\! 10^{-3}$ & $K^{\alpha}_d$\\
\hline
1.075  & 55933 &  2.017 & 1.562 & 2.0004 & 73774 &  3.638 & 718.7 & -0.0020 \\
1.100  & 53971 &  2.035 & 1.520 & 1.9996 & 71792 &  3.563 & 679.3 & -0.0022 \\
1.150  & 50311 &  2.057 & 1.446 & 1.9993 & 66465 & -2.545 & 9.926 & 1.9252 \\
1.200  & 46985 &  2.063 & 1.387 & 1.9995 & 58971 & -2.561 & 2.460 & 1.9769 \\
1.250  & 43949 &  2.056 & 1.339 & 1.9993 & 52256 & -2.591 & 1.433 & 1.9889 \\
1.300  & 41159 &  2.035 & 1.298 & 1.9968 & 46230 & -2.556 & 1.030 & 2.0000 \\
1.350  & 38569 &  1.993 & 1.262 & 1.9922 & 40821 & -2.391 & 0.820 & 2.0261 \\
1.375  & 37338 &  1.937 & 1.239 & 1.9757 & 38329 & -2.116 & 0.764 & 2.0785 \\
1.400  & 35965 & -0.801 & 0.491 & 1.4647 & 36149 &  11.59 & 1.327 & 2.0813 \\
\hline
\end{tabular}
\end{table*}

\begin{table}
\caption{Comparison of the present theoretical and available experimental radiative lifetimes (in milliseconds) for the $a^3\Pi(\Omega^{\pm},v^{\prime},J^{\prime})$ rovibronic levels of the $^{12}$C$^{16}$O isotopomer. $^a$ - Ref.~\cite{Fournier}, $^b$ - Ref.~\cite{tauCO2000}, $^c$ - Ref.~\cite{tauCO1999}, $^d$ - Ref.~\cite{tauCO2007}.}\label{Tabtau}
\begin{tabularx}{\columnwidth}{p{0.05\columnwidth}p{0.05\columnwidth}p{0.05\columnwidth}p{0.15\columnwidth}p{0.15\columnwidth}p{0.15\columnwidth}p{0.25\columnwidth}}
\hline\hline
$\Omega^{\pm}$ & $v^{\prime}$ & $J^{\prime}$ & \multicolumn{3}{c}{Theory} & Experiment \\
&    &    &   FS-RCC   &   RCI   & SSI  \\
\hline
0$^+$ & 0 & 0 & 690 & 640 & 285 &  90$\pm$2~$^a$\\
0$^+$ & 3 & 3 &  62 &  58 &  66 &  73$\pm$22~$^b$\\
\hline
0$^+$ & 3 & 2 & 102 &  84 & 104 & 119$\pm$36~$^b$\\
0$^-$ & 3 & 2 & 147 & 170 & 165 & 147$\pm$44~$^b$\\
\hline
1$^+$ & 0 & 1 & 2.59 & 3.02 & 2.91 & 2.63$\pm$0.03~$^d$\\
1$^+$ & 3 & 1 & 2.67 & 3.15 & 3.02 & 3.46$\pm$0.33~$^b$\\
1$^-$ & 3 & 2 & 2.75 & 3.24 & 3.11 & 3.04$\pm$0.38~$^c$\\
      &   &   &      &      &      & 3.67$\pm$0.33~$^b$\\
\hline
2$^+$ & 0 & 2 & 142 & 165 & 159 & 143$\pm$4~$^d$\\
2$^+$ & 3 & 2 & 154 & 179 & 173 & 211$\pm$63~$^b$\\
2$^+$ & 3 & 3 &  65 &  75 &  73 &  72$\pm$22~$^b$\\
\hline
\end{tabularx}
\end{table}

\section{Concluding remarks}
The transition probabilities of the intercombination Cameron system of carbon monoxide have been computationally studied in the framework of three different \emph{ab initio} methods.
The most reliable spin-forbidden $a^3\Pi_{\Omega=0^+,1} - X^1\Sigma^+$ transition dipole moments were obtained using the multi-reference Fock space coupled cluster method which was combined with the empty-core generalized relativistic pseudopotential model to introduce proper treatment of relativistic effects (including Breit interactions) into all-electron correlation calculations.

The radiative lifetimes evaluated for the particular rovibronic levels of the fine structure $a^3\Pi_{\Omega=0^{\pm},1,2}$ components are in a very good agreement with their most accurate experimental counterparts. The sensitivity coefficients of the CO Cameron system to a presumable drift of the fine structure constant have demonstrated very smooth $R$-dependence except the narrow region of the local spin-orbit coupling. It means that the future deperturbation analysis of even very small $a^3\Pi\sim a'^3\Sigma^+$ perturbations can provide much higher sensitivity coefficients of the CO Cameron bands.

\section*{Acknowledgements}

The work was supported by the Russian Science Foundation (RSF), Grant No.18-13-00269. The work on the GRPP generation for the light elements was supported by the personal scientific fellowship for N.S.\ Mosyagin from the governor of the Leningrad district.

\bibliography{co}

\end{document}